\documentclass[twocolumn,preprintnumbers,amsmath,amssymb,superscriptaddress]{revtex4}
\usepackage{graphicx}% Include figure files
\usepackage{dcolumn}% Align table columns on decimal point
\usepackage{bm}% bold math

\newcommand{\bea}{\begin{equation}}
\newcommand{\eea}{\end{equation}}

\begin{document}

\title{Slippage of Newtonian liquids: influence on the dynamics of dewetting thin films}

\author{R. Fetzer, K. Jacobs\thanks{To whom correspondence should be addressed. E-mail: k.jacobs@physik.uni-saarland.de.}\\
{\it Department of Experimental Physics, Saarland University,}\\
{\it D-66041 Saarbr\"ucken, Germany}}
% \email{r.fetzer@physik.uni-saarland.de}
% \email{k.jacobs@physik.uni-saarland.de}

\date{\today}

%%%%%%%%%%%%%%%%%%%%%%%%%%%%%%%%%%%%%%%%%%%%%%%%%%%%%%%%%%%%%%%%%%%%%%%%%%%
\begin{abstract}
Slippage of Newtonian liquids in the presence of a solid substrate
is a newly found phenomenon the origin of which is still under
debate. In this paper, we present a new analysis method to extract
the slip length. Enhancing the slip of liquids is an important issue
for microfluidic devices that demand for high throughput at low
pumping power. We study the velocity of short-chained liquid
polystyrene (PS) films dewetting from non-wettable solid substrates.
We show how the dynamics of dewetting is influenced by slippage and
we compare the results of two types of substrates that give rise to
different slip lengths. As substrates, Si wafers were used that have
been coated by octadecyltrichlorosilane (OTS) or
dodecyltrichlorosilane (DTS), respectively. Our results demonstrate
that the dewetting velocity for PS films on DTS is significantly
larger than on OTS and that this difference originates from the
different slip lengths of the liquid on top of the two surfaces. For
PS films of thicknesses between 130~nm and 230~nm we find slip
lengths between 400~nm and 600~nm, depending on substrate and
temperature.

\end{abstract}

\maketitle

%PACS: 68.15.+e, 47.20.Ma, 47.54.+r, 68.37.Ps
                  % Valid PACS numbers may be entered using the
                  % \verb+\pacs{#1}+ command.

\section{Introduction}

 Thin films of liquid polymers play an important role in
numerous technological processes ranging from lithography to
biological membranes \cite{Oro97}. A key to understanding the
stability of, e.g., liquid coatings on solid surfaces was the
calculation \cite{Vri66,Ruc74,Sha85,Die88,Sch89} and the
experimental derivation of the effective interface potential
\cite{Rei92,Sfe98,Kim99,See01,See01-2,See01-3}. In recent years, one
major focus of interest has been to understand the dynamics and
morphology of polymer films dewetting from non-wettable substrates
\cite{Red91,Red94,Bro94,Bro97,Mas02,Dam03,Rei01,Vil06,Jac98,Rei00,See01-4,Bec03,Net03,Fet05,Vil05}.
The dewetting rate of the liquid hereby strongly depends on the flow
velocity at the solid/liquid interface. Hitherto it was common sense
that the occurrence of a non-zero flow velocity (slippage) at the
interface is restricted to high molecular weight polymer films
\cite{Gen85}. Only recently, new experimental techniques have
revealed that also Newtonian liquids may exhibit slippage
\cite{Pit00,Sch05,Cot02,Leg03}. The occurrence and the nature of
slippage is of large technological interest, since a sliding fluid
can flow faster through, e.g., microfluidic devices.

Many studies have focused on techniques to measure slippage. These
approaches can be classified in direct and indirect methods to
determine the fluid velocity profile near the solid/liquid
interface. Direct methods use, e.g., tracer particles in
combination with near-field laser velocimetry \cite{Leg03,Herv03}
or fluorescence recovery after photobleaching techniques
\cite{Pit00,Sch05,Mig93}. Indirect methods aim at measuring the
drainage force of an object moved in a liquid to calculate the
amount of slippage. Common techniques use a surface forces
apparatus \cite{Cot02,Zhu02} or an atomic force microscope (AFM)
with a colloidal probe \cite{Cra01}. For details see, e.g., the
review articles from Lauga et al. \cite{Lau05} or Neto et al.
\cite{Net05}.

A common measure of slippage is the slip length, which can be
understood as the length below the solid/liquid interface where
the velocity extrapolates to zero. For simple fluids, the slip
length is found to be independent of the shear rate
\cite{Herv03,Pri04}, but is influenced by the wettability of the
substrate as well as by the roughness of the wall: Molecular
dynamics simulations for a simple liquid with 140$^\circ$ contact
angle on the surface of interest found a slip length in the order
of 30 diameters of the fluid molecules, but no-slip boundary
condition for a wetting situation \cite{Bar99}. This result is in
agreement with experimental data of hexadecane
\cite{Pit00,Sch05,Herv03} or glycerol \cite{Cot02}. The role of
surface roughness is ambiguous: simulations revealed that,
depending on the pressure, surface roughness can increase or
decrease the amount of slippage \cite{Cot03}; experiments
demonstrated the importance of the lateral length scale of the
roughness \cite{Leg03,Jos06}.

In our studies, we investigate the dewetting process of thin
liquid films, a typical scenario is depicted in
Fig.~\ref{fig_optisch}. In order to extract the slip length, we
develop a new method for the analysis of the dewetting rates.
Dewetting can be described in three different stages \cite{Bro97},
starting from the very beginning of hole formation. In the first
stage, the dewetted region grows exponentially in time and, in the
vicinity of the dry area, a homogeneous thickening of the film can
be observed; the hole does not yet exhibit a rim. In the second
regime, the 'birth' of the rim takes place. In this stage, the rim
has an asymmetric shape, and the radius grows linearly in time. In
the third stage of dewetting, which starts at radii in the order
of the slip length, surface tension rounds the rim which becomes
more symmetric and grows from now on in a self-affine manner. Many
experiments confirm this three-stage dewetting scenario
\cite{Mas02,Dam03,Rei01,Vil06}.

In the third stage of a 'mature' rim, slippage is found to
influence the dynamics in a substantial way \cite{Red94,Bro94}:
For the case of a no-slip boundary condition or negligible sliding
friction at the solid/liquid interface, viscous dissipation (which
occurs dominantly in the vicinity of the three phase contact line)
results in a linear growth of holes; if, however, dissipation at
the solid/liquid interface dominates the dewetting process, an
$R\propto t^{2/3}$ power law is expected. For the latter case of
large slippage, the third stage might be followed by a forth
regime, when the rim height is large as compared to the slip
length \cite{Bro94}. In this case, sliding dissipation again can
be neglected and a linear growth $R\propto t$ is expected.

In our studies we are interested in the stage of a mature rim that
grows in a self-affine manner in width and height. In the
following we will demonstrate how we can deduce the slip length
from the dynamics of the hole growth. We hereby restrict ourselves
to liquids of Newtonian behavior and therefore investigate only
the dewetting dynamics of short-chained polymer melts below the
entanglement length. A short glance on Fig.~\ref{fig_r_t} reveals
that the growth law for the hole radius is clearly non-linear. We
therefore suppose that slippage cannot be neglected. On the other
hand, the polymer melt is not entangled. Hence, slippage is not
expected to dominate the growth of holes at any stage. In the
theoretical section we will describe a model assuming that neither
viscous dissipation at the contact line nor sliding friction at
the solid/liquid interface entirely dominates the dynamics. We
rather propose that the two different dissipation mechanisms are
balanced and both have to be taken into account simultaneously
\cite{Jac98}.

\section{Experimental Section}

For the experiments, atactic polystyrene (PS, by Polymer Standard
Service, Mainz, Germany) with a molecular weight of 13.7 kg/mol
and a polydispersity of $M_w/M_n=1.03$ was used. The samples were
prepared by spin casting a toluene solution of PS onto mica,
floating the films on Millipore$^{TM}$ water, and then
transferring them to hydrophobic substrates. For the
hydrophobization, we used two different self-assembled monolayer
coatings on top of Si wafers (2.1~nm native oxide layer),
octadecyltrichlorosilane (OTS) and the shorter
dodecyltrichlorosilane (DTS), respectively, prepared by standard
techniques \cite{Was89}. By ellipsometry (EP$^3$ by Nanofilm,
G\"ottingen, Germany), the thickness of the SAMs was found to be
$d_{OTS}=2.4$~nm and $d_{DTS}=1.5$~nm, respectively. Surface
characterization by atomic force microscopy (AFM, Multimode by
Veeco, Santa Barbara, USA) revealed an RMS roughness of 0.09(1)~nm
(OTS) and 0.13(2)~nm (DTS) at a $(1~\mu$m$)^2$ scan size, and a
static contact angle of polystyrene droplets of
$\theta_Y=67(3)^\circ$ on both coatings. The surface tension of
polystyrene is $\gamma=30.8$~mN/m.

\begin{figure}\centering
\includegraphics*[width=\linewidth]{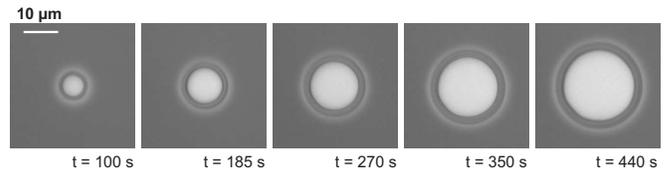}
%\centerline{\epsfig{file=speeds.eps, width=9.5cm}}
\caption{Growth of a hole in a 130 nm thick PS(13.7k) film on DTS,
captured by optical microscopy at $120\,^\circ$C.}
\label{fig_optisch}
\end{figure}

\begin{figure}\centering
\includegraphics*[width=6.5cm]{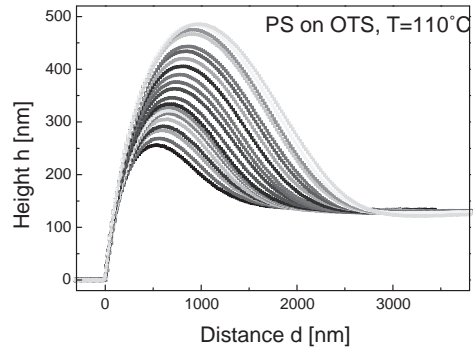}
%\centerline{\epsfig{file=speeds.eps, width=9.5cm}}
\caption{Snapshots of the rim profile of a dewetting PS(13.7k)
film on OTS, measured in situ with tapping mode AFM. The dynamic
contact angle stays constant during the growth of the hole. Here,
the three phase contact line is shifted to the origin.}
\label{fig_h_r_dyn}
\end{figure}

The polystyrene films in this study are either 130(5)~nm or
230(5)~nm thick. To induce dewetting, the films were heated above
the glass transition temperature of PS ($93\,^\circ$C) to three
different temperatures ($110\,^\circ$C, $120\,^\circ$C,
$130\,^\circ$C). After a few seconds, circular holes are born and
instantly start to grow. An example of a typical temporal series
captured by optical microscopy is shown in Fig.~\ref{fig_optisch}.
Due to mass conservation of the liquid rims develop, surrounding
each hole. Cross sections of such a growing rim are shown in
Fig.~\ref{fig_h_r_dyn}, as measured by in situ AFM. The dynamic
contact angle of the liquid is $\theta_{dyn}=56(2)^\circ$ and
stays constant during dewetting.

\begin{figure}\centering
\includegraphics*[width=7.5cm]{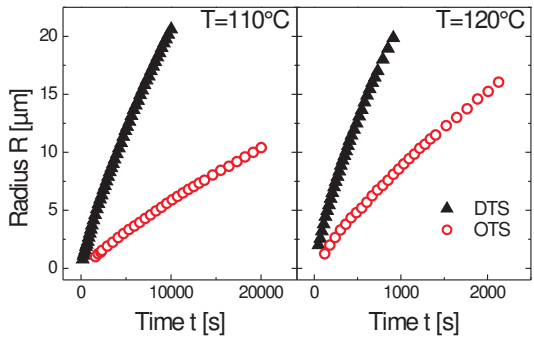}
%\centerline{\epsfig{file=speeds.eps, width=9.5cm}}
\caption{Radius of dewetted region as a function of time, measured
for 130 nm thick PS(13.7k) films on OTS as well as on DTS
substrates at two different temperatures.} \label{fig_r_t}
\end{figure}

\begin{figure}\centering
\includegraphics*[width=6.5cm]{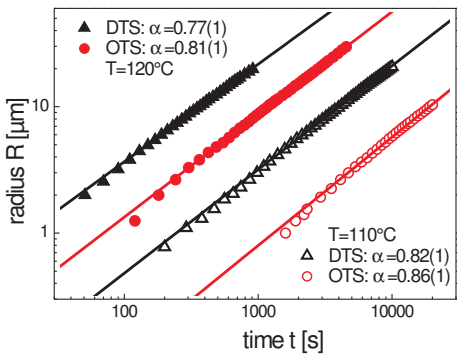}
%\centerline{\epsfig{file=speeds.eps, width=9.5cm}}
\caption{Data shown in Fig.~\ref{fig_r_t} in logarithmic scale.
For different dewetting temperatures and on the different
substrates, the radii grow algebraically in time with an exponent
of $\alpha=0.82(5)$.} \label{fig_r_t_log}
\end{figure}

In Fig.~\ref{fig_r_t}, the optically measured radii of the
emerging holes in the PS(13.7k) films as a function of time are
depicted for two temperatures and the two types of silane
substrate. Firstly, for each series, we find a non-linear growth
law. To emphasize this, the data are additionally drawn in
Fig.~\ref{fig_r_t_log} in a logarithmic diagram, revealing an
algebraic growth of the radius in time with an exponent of
$\alpha=0.82(5)$. Secondly, dewetting progresses clearly faster on
DTS than on OTS coated substrates.

We qualitatively interpret this observation as follows. At the
same temperature the polymer films on both samples have identical
properties: the viscosity $\eta$ as well as the surface tension
$\gamma$ do not depend on the substrate underneath. Additionally,
the static contact angle $\theta_Y$ of polystyrene on both
surfaces is constant within the experimental error. Therefore, the
spreading coefficient $S=\gamma (\cos\theta_Y-1)= -0.0188$~N/m
which is the driving force of the dewetting process is identical
on OTS and on DTS substrates. The only parameter that could be
different is the boundary condition at the solid/liquid interface
of the two substrate. Since dewetting on DTS is much faster than
on OTS, we expect a significantly larger slip length for PS(13.7k)
on the DTS coating.

\section{Theoretical Section: Energy dissipation models}

To get more insight in the mechanisms involved in the dewetting
process, a detailed analysis of the velocity is required. The
velocity of the moving front $V=\dot{R}$ is given by the force
balance between the driving force $|S|$ and the power $P$
dissipated in the dewetting process, $|S|\cdot V=P$. Assuming no
slip at the solid/liquid interface, energy is dissipated by
viscous friction only within the liquid volume and at the three
phase contact line. Since the highest shear rates arise in the
vicinity of the contact line, viscous dissipation $P_v$ takes
place dominantly at this line and is therefore independent of the
size of the rim. However, $P_v$ is influenced by the flow geometry
near the contact line, i.e., by the dynamic contact angle
$\theta_{dyn}$. Thus, the dewetting velocity $V=v_v$ is given by
\bea v_v=C_v(\theta_{dyn})\,\frac{|S|}{\eta}\label{eq_vv}\eea with
$\eta$ the viscosity of the liquid, and $C_v(\theta_{dyn})$ a
function of the dynamic contact angle that counts for the flow
geometry \cite{Bro94}.

As shown in Fig.~\ref{fig_h_r_dyn}, the dynamic contact angle is
temporally constant in our experiments. Thus, viscous dissipation
does neither depend on hole size nor on time. On the other hand,
the driving force $|S|$ does not vary during the dewetting
process. Hence, in the model of no-slip boundary condition, the
dewetting velocity $V=v_v$ is constant, and the radius $R$ grows
linearly in time, $R\propto t$, which holds for Newtonian liquids
as shown, e.g., in Ref.~\cite{Red91}.

Since the data in Fig.~\ref{fig_r_t} clearly do not show a linear
growth of the radius $R$, but rather a decreasing velocity, our
experiments do not allow the assumption of a no-slip boundary
condition.

Another model, the so-called 'full-slip' or 'plug flow' model,
introduced to explain the dewetting rate of entangled polymer
melts, assumes large slip lengths at the solid/liquid interface
\cite{Bro94}. Here, energy dissipation dominantly occurs at this
interface over the distance of about the width of the rim, $w$.
Hence, the power dissipated in the dewetting process is
proportional to $w$, and force balance results in the velocity
$V=v_s$, \bea
v_s=\frac{1}{3}\frac{|S|}{\eta}\frac{b}{w}\;,\label{eq_vs}\eea
with the slip length $b$. Self-similarity of the growing mature
rim, (which means a constant dynamic contact angle, c.f.
Fig.~\ref{fig_h_r_dyn}), and mass conservation of the liquid
yields $w\propto\sqrt{R}$. Hence, in this model, the velocity
decreases with increasing hole size, $v_s\propto 1/\sqrt{R}$, and
the hole radius increases algebraically in time with an exponent
$\alpha=2/3$, $R\propto t^{2/3}$. For simplicity, we can introduce
the constant number $K$ and write $v_s=K/\sqrt{R}$.

Although the data in Fig.~\ref{fig_r_t_log} show an algebraic
behavior $R\propto t^{\alpha}$, the exponents fitted to the data
achieve values clearly above 2/3; the mean value of the exponents
of the four curves shown in Fig.~\ref{fig_r_t_log} is
$\alpha=0.82(5)$. Hence, the model of full slippage also is not
adequate to describe our experimental results.

Another model combining energy dissipation at the three phase
contact line and sliding dissipation at the solid/liquid interface
is required. In literature there are some suggestions of a
transition within the dewetting process, meaning that at an early
stage viscous dissipation at the contact line dominates the
dynamics of dewetting, $R\propto t$, while in a later stage the
holes grow mostly due to slip effects, $R\propto t^{2/3}$
\cite{Bro97,Mas02,Dam03}. In our experiments, for some small
holes, we can see such a transition from an early stage of almost
constant dewetting velocity to a later stage of decreasing
velocity. Nevertheless, the model of slip-dominated dissipation is
not appropriate for the experimental data, even not at the later
stage.

In our earlier study, Ref.~\cite{Jac98}, simple addition of both
contributions of the dewetting velocity is suggested ('combined
model'), \bea V=v_v+v_s \label{eq_comb}\eea with $v_v=const.$ and
$v_s=K/\sqrt{R}$. Separation of the variables yields the function
\begin{eqnarray}
t-t_0&=&\frac{1}{v_v}\left(R-2\gamma\sqrt{R}+2\gamma^2\ln\left(1
+\frac{\sqrt{R}}{\gamma}\right)\right)\\
\gamma&=&K/v_v\; .\nonumber\end{eqnarray}
Unfortunately, this
implicit equation for the radius is rather cumbersome. Since it can
be fitted to any data of hole growth, this function is not able to
critically test the combined model and its assumptions.

%\bea
%t-t_0&=&\frac{1}{v_v}\left(R-2\gamma\sqrt{R}+2\gamma^2\ln\left(1
%+\frac{\sqrt{R}}{\gamma}\right)\right)\nn\\\gamma&=&K/v_v\; .\eea

\section{Testing the combined model}

\begin{figure}\centering
\includegraphics*[width=7.5cm]{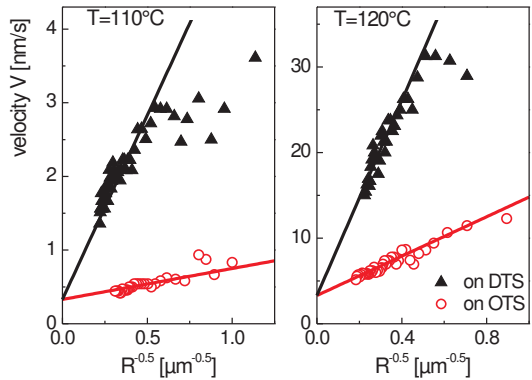}
%\centerline{\epsfig{file=speeds.eps, width=9.5cm}}
\caption{Except for very small holes, the data of the dewetting
velocity $V=\dot{R}$ are in line when plotting versus
$1/\sqrt{R}$. Extrapolation for infinitely large holes yields to a
finite velocity $v_0$.} \label{fig_v_sqrtR}
\end{figure}

An alternative and more comprehensive way to check the combined
model Eq.~(\ref{eq_comb}) is to plot the velocity versus
$1/\sqrt{R}$. If the combined model is valid, the data points
should then appear on a straight line with intercept $v_v$ and
slope $K$. As shown in Fig.~\ref{fig_v_sqrtR}, our experimental
data indeed lie on such a straight line with a finite velocity
$v_0$ extrapolating $1/\sqrt{R}$ to zero, i.e., for infinitely
large holes. This gives a first hint that the simple 'combined
model' $V=v_v+v_s$ is an adequate description for the dewetting
rate of short chain polymer films. Just the data of small holes on
DTS with radii below 4~$\mu$m deviate from this line. Since the
rims of these holes are not yet in the mature regime, we can
neglect these data points for further analysis.

\begin{figure}\centering
\includegraphics*[width=8.5cm]{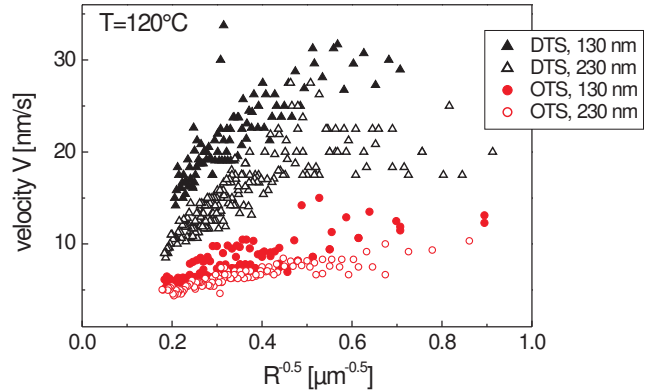}
%\centerline{\epsfig{file=speeds.eps, width=9.5cm}}
\caption{For evaluation of the combined model all measured data of
dewetting velocities of PS(13.7k) films at 120$\,^\circ$C are
plotted versus $1/\sqrt{R}$.} \label{fig_alle_at120}
\end{figure}

How can we further test that the combined model is indeed an
appropriate description?

If the intercept $v_0$ represents the dewetting velocity of
viscous flow, $v_v$, then it should fulfill three conditions: i)
the temperature dependence of $v_0$ should be dominated by the
inverse viscosity of the polymer melt, ii) $v_0$ should be
independent of the initial film thickness $h_0$, which is a very
strong condition, and iii) the intercept $v_0$ should only depend
on the contact angle, but not on the slip length.

\begin{figure}\centering
\includegraphics*[width=8cm]{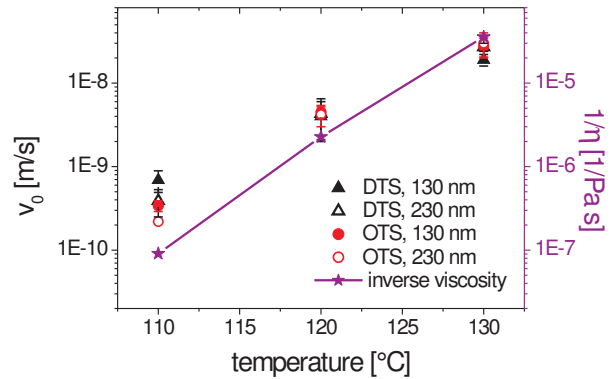}
%\centerline{\epsfig{file=speeds.eps, width=9.5cm}}
\caption{The finite velocity $v_0$ extrapolated for infinite large
hole radii increases with increasing dewetting temperature. This
behavior qualitatively compares well to the temperature dependence
of the inverse melt viscosity. Additionally, $v_0$ does neither
depend on the initial film thickness nor on the substrate
coating.} \label{fig_A_eta}
\end{figure}

To test these three conditions, we repeated our dewetting
experiments at three different temperatures and with films of two
different thicknesses on both coatings DTS and OTS, cf.
Fig.~\ref{fig_alle_at120} for the experiments at 120$\,^\circ$C.
As shown in Fig.~\ref{fig_A_eta}, the extrapolated velocity $v_0$
increases for increasing dewetting temperature, and, moreover, the
qualitative run compares well to the inverse viscosity of
PS(13.7k) as measured by a rheometer. Note that viscous
dissipation in the vicinity of the three phase contact line is not
necessarily expected to exactly follow the $1/\eta$ behavior,
since there is a contribution additionally to the shear flow that
derives directly from the contact line dynamics, i.e., thermally
activated jumps over an energy barrier, pinning, etc.
\cite{Rui99}. Thus, the first condition for $v_0$ concerning the
temperature dependence is satisfied sufficiently. From
Fig.~\ref{fig_A_eta}  we also learn that the initial film
thickness $h_0$ does not systematically influence the extrapolated
velocity $v_0$, although the measured velocity at finite hole size
does depend on film thickness, cf. Fig.~\ref{fig_alle_at120}. With
this, the second condition is fulfilled. Since the intercept $v_0$
is independent of initial film thickness, it also does not depend
on the width of the rim and, consequently, not on the effect of
slippage at the solid/liquid interface. Additionally, $v_0$ is not
influenced by the substrate, OTS or DTS covered silicon wafers, on
which the PS films are supposed to show different slip lengths.
This satisfies the third condition, the independence of slippage
of $v_0$. Note that in general $v_0$ depends on the substrate,
since viscous dissipation is determined by the dynamic contact
angle $\theta_{dyn}$. In our experiments, however, we found the
same $\theta_{dyn}$ on both substrates and, consequently, same
values for $v_0$. With the described tests we found all three
conditions satisfied and therefore can assume that the intercept
$v_0$ indeed compares to $v_v$, which allows further analysis of
the data.

\section{Results: The slip length}

\begin{figure}\centering
\includegraphics*[width=7cm]{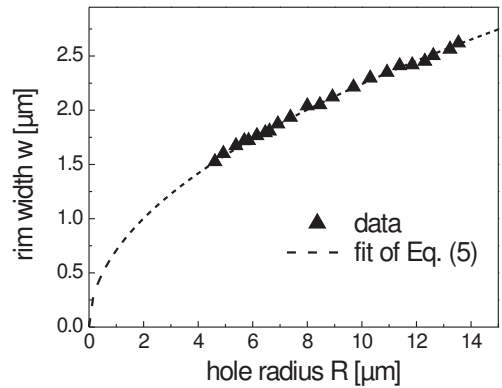}
%\centerline{\epsfig{file=speeds.eps, width=9.5cm}}
\caption{Width of the rim as a function of the hole radius,
exemplarily shown for a 130~nm thick PS(13.7k) film dewetting from
OTS at 110$\,^\circ$C. By fitting Eq.~(\ref{eq_w}) to the data we
get $C_s=1.96$.} \label{fig_w_R}
\end{figure}

In a second step the velocity component $v_s$ is analyzed and the
slip length $b$ is extracted. In the case of full slippage at the
solid/liquid interface the velocity is given by Eq.~(\ref{eq_vs}),
where the width of the rim reads \bea w=C_s\cdot\sqrt{h_0
R}\;.\label{eq_w}\eea Hence, the slope $K$ shown in
Fig.~\ref{fig_v_sqrtR} is expected to be \bea
K=\frac{1}{3}\frac{|S|}{\eta}\frac{b}{C_s\sqrt{h_0}}\;.\label{eq_K_b}\eea
First, the validity of Eq.~(\ref{eq_w}) is tested by plotting the
width of the rim, $w$,  versus the hole radius $R$, cf.
Fig.~\ref{fig_w_R}. Here, $w$ is taken as the lateral distance
between the three phase contact line ($x = 0$) and the position
where the rim height is dropped to 110\% of the prepared film
thickness $h_0$, i.e., $h(w)=1.1\,h_0$. Fitting Eq.~(\ref{eq_w})
to the data of 130~nm and 230~nm thick PS films, we obtain
$C_s=1.96(5)$ on OTS and $C_s=2.1(1)$ on DTS covered wafers.

\begin{figure}\centering
\includegraphics*[width=9cm]{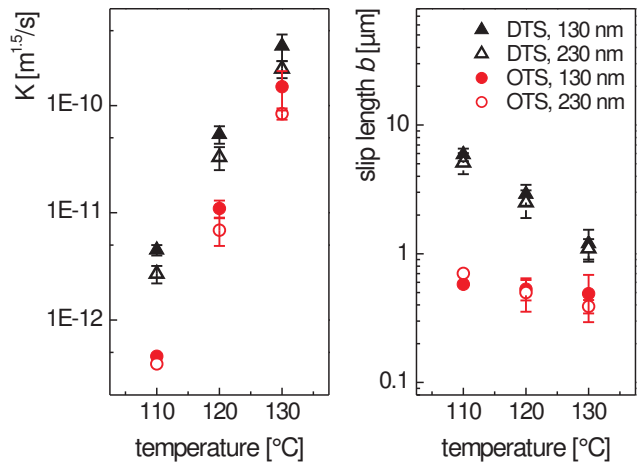}
%\centerline{\epsfig{file=speeds.eps, width=9.5cm}}
\caption{Left: Slope $K$ of the straight lines in
Fig.~\ref{fig_v_sqrtR} as a function of the melt temperature. The
parameter $K$ depends on the initial film thickness as well as on
the type of coating. Right: The slip length for polystyrene on top
of both DTS and OTS covered Si wafers. The values are determined
from $K$.} \label{fig_B_eta}
\end{figure}

Knowing $C_s$, the spreading coefficient $|S|=0.0188$~N/m, the
initial film thickness $h_0$, and the viscosity $\eta$ of the
investigated films, cf. Fig.~\ref{fig_A_eta}, we can now use Eq.
(\ref{eq_K_b}) to directly determine the slip length $b$ from the
slope $K$. Although the values for $K$ are affected by the film
thickness,  cf. Figs.~\ref{fig_alle_at120} or \ref{fig_B_eta}a,
this dependence is canceled out very well for the slip length, as
shown in Fig.~\ref{fig_B_eta}b. This again indicates that the
combined model Eq.~(\ref{eq_vs}) can be used to analyze the data
of our experimental system. We find $b$ decreasing for increasing
dewetting temperature. Additionally, the slip length of PS(13.7k)
is clearly larger on DTS than on OTS, but the difference decreases
with increasing temperature, a fact that corroborates the results
for the slip lengths determined by rim shape analysis
\cite{Fet06}.

\section{Conclusion}

We have developed a new analysis method to extract slip lengths
from dewetting rates. The experiments give evidence that a model
has to be put forward that combines frictional energy dissipation
at the interface with viscous dissipation at the three phase
contact line. We show that simple addition of two different
velocity components is an adequate description for highly viscous
Newtonian liquids dewetting from smooth surfaces. We moreover were
able to demonstrate that this description also captures a variable
friction coefficient at the solid/liquid interface. The components
$v_v$ and $v_s$ are given by the well known expressions derived
for the limiting cases no-slip and plug flow, respectively. From
the slope of velocity data plotted versus $1/\sqrt{R}$ the slip
length can be determined.

For short-chained polystyrene melts on OTS and DTS covered wafers
we found slip lengths between 400~nm and 6 $\mu$m. Slippage in
this system decreases for increasing dewetting temperature. On the
DTS coating the slip lengths are up to one order of magnitude
larger than on OTS.

So far, we have only hypotheses for the molecular mechanisms at
the solid/liquid interface that give rise to the different slip
lengths on OTS and DTS covered wafers. The commonly investigated
system parameters like interaction forces between solid and
liquid, i.e., contact angle and long-ranged dispersion forces, or
substrate roughness are identical on both types of substrate.
Thus, we have no straightforward explanation. The molecular origin
of slippage in our system has to be postponed to future research.
The simple analysis method of dewetting rates, however, is a
powerful tool that allows for extensive studies of various systems
to get a comprehensive picture of slippage of simple and complex
liquids.

\section{Acknowledgments}

We acknowledge financial support by the European Graduate School
GRK 532 and by grant JA~905/3 within the DFG priority program
1164, and generous support of Si wafers by Siltronic AG,
Burghausen, Germany.

%\clearpage

%\begin{figure}\centering
%\includegraphics*[width=\linewidth]{TOCgraphic.eps}
%\centerline{\epsfig{file=speeds.eps, width=9.5cm}}
%\caption{TOC graphic}
%\end{figure}


\begin{thebibliography}{0}

\bibitem{Oro97}
Oron, A.; Davis, S. H.; Bankoff, S. G. {\it Rev. Mod. Phys.} {\bf
1997}, {\it 69}, 931.

\bibitem{Vri66}Vrij, A. {\it Discuss. Faraday Soc.} {\bf 1966}, {\it
42}, 23.

\bibitem{Ruc74}Ruckenstein, E.; Jain, R. K. {\it J. Chem. Soc. Faraday Trans.} {\bf 1974}, {\it
II}, 132.

\bibitem{Sha85}Sharma, A.; Ruckenstein, E. {\it J. Colloid Interface
Sci.} {\bf 1985}, {\it 106}, 12.

\bibitem{Die88}Dietrich, S. In {\it Phase Transition and Critical
Phenomena}; Domb, C., Lebowitz, J. L., Eds.; Academic: London,
1988; Vol. 12.

\bibitem{Sch89}Schick, M. In {\it Liquids at Interfaces};
Charvolin, J., Joanny, J. F., Zinn-Justin, J., Eds.; Elsevier
Science: Amsterdam, 1989.

\bibitem{Rei92}Reiter, G. {\it Phys. Rev. Lett.} {\bf 1992}, {\it
68}, 75.

\bibitem{Sfe98}Sferrazza, M.; Heppenstall-Butler,  M.; Cubitt, R.; Bucknall, D.; Webster, J.; Jones, R. A. L. {\it Phys. Rev. Lett.} {\bf
1998}, {\it 81}, 5173.

\bibitem{Kim99}Kim, H. I.; Mate, C. M.; Hannibal, K. A.; Perry, S. S. {\it Phys. Rev. Lett.} {\bf 1999},
{\it 82}, 3496.

\bibitem{See01}Seemann, R.; Herminghaus, S.; Jacobs, K. {\it Phys. Rev.
Lett.} {\bf 2001}, {\it 86}, 5534.




\bibitem{See01-2}Seemann, R.; Blossey, R.; Jacobs, K. {\it J. Phys.: Condens.
Matter} {\bf 2001}, {\it 13}, 4915.

\bibitem{See01-3}Seemann, R.; Herminghaus, S.; Jacobs, K. {\it J. Phys.: Condens.
Matter} {\bf 2001}, {\it 13}, 4925.


\bibitem{Red91}Redon, C.; Brochard-Wyart, F.; Rondelez, F. {\it Phys. Rev.
Lett.} {\bf 1991}, {\it 66}, 715.

\bibitem{Red94}Redon, C.; Brzoska, J. B.; Brochard-Wyart, F. {\it Macromolecules} {\bf 1994}, {\it 27}, 468.

\bibitem{Bro94}Brochard-Wyart, F.; de Gennes, P.-G.; Hervet, H.; Redon, C. {\it
Langmuir} {\bf 1994}, {\it 10}, 1566.

\bibitem{Bro97}Brochard-Wyart, F.; Debregeas, G.; Fondecave, R.; Martin, P. {\it
Macromolecules} {\bf 1997}, {\it 30}, 1211.

\bibitem{Mas02}Masson, J.-L.; Green, P. F. {\it Phys. Rev. Lett.} {\bf 2002}, {\it 88},
205504.

\bibitem{Dam03}Damman, P.; Baudelet, N.; Reiter, G. {\it Phys. Rev. Lett.} {\bf
2003}, {\it 91}, 216101.

\bibitem{Rei01}Reiter, G. {\it Phys. Rev. Lett.} {\bf 2001}, {\it 87},
186101.

\bibitem{Vil06}Vilmin, T.; Rapha\"{e}l, E.; Damman, P.;
Sclavons, S.; Gabriele, S.; Hamieh, M.; Reiter, G. {\it Europhys.
Lett.} {\bf 2006}, {\it 73}, 906.

\bibitem{Jac98}Jacobs, K.; Seemann, R.; Schatz, G.; Herminghaus, S. {\it
Langmuir} {\bf 1998}, {\it 14}, 4961.


\bibitem{Rei00}Reiter, G.; Khanna, R. {\it Langmuir} {\bf 2000}, {\it 16}, 6351.

\bibitem{See01-4}Seemann, R.; Herminghaus, S.; Jacobs, K. {\it Phys. Rev.
Lett.} {\bf 2001}, {\it 87}, 196101.




\bibitem{Bec03}Becker, J.; Gr\"un, G.; Seemann, R.; Mantz, H.;
Jacobs, K.; Mecke, K. R.; Blossey, R. {\it Nature Materials} {\bf
2003} {\it 2}, 59.

\bibitem{Net03}Neto, C.; Jacobs, K.; Seemann, R.; Blossey, R.;
Becker, J.; Gr\"un, G. {\it J. Phys.: Condens. Matter} {\bf 2003},
{\it 15}, 3355.

\bibitem{Fet05}Fetzer, R.; Jacobs, K.; M\"unch, A.; Wagner, B.;
Witelski, T. P. {\it Phys. Rev. Lett.} {\bf 2005}, 95, 127801.

\bibitem{Vil05}Vilmin, T.; Rapha\"{e}l, E. {\it Europhys.
Lett.} {\bf 2005}, {\it 72}, 781.


\bibitem{Gen85}de Gennes, P. G. {\it Rev. Mod. Phys.} {\bf 1985}, {\it 57}, 3.

\bibitem{Pit00}Pit, R.; Hervet, H.; L\'{e}ger, L. {\it Phys. Rev.
Lett.} {\bf 2000}, {\it 85}, 980.

\bibitem{Sch05}Schmatko, T.; Hervet, H.; Leger, L. {\it Phys. Rev.
Lett.} {\bf 2005}, {\it 94}, 244501.




\bibitem{Cot02}Cottin-Bizonne, C.; Jurine, S.; Baudry, J.; Crassous, J.; Restagno, F.; Charlaix, E. {\it Eur. Phys. J.
E} {\bf 2002}, {\it 9}, 47.

\bibitem{Leg03}
L\'eger, L. {\it J. Phys.: Condens. Matter} {\bf 2003}, {\it 15},
S19.

\bibitem{Herv03}Hervet, H.; L\'{e}ger, L. {\it C. R.
Physique} {\bf 2003}, {\it 4}, 241.

\bibitem{Mig93}Migler, K. B.; Hervet, H.; L\'{e}ger, L. {\it Phys. Rev.
Lett.} {\bf 1993}, {\it 70}, 287.

\bibitem{Zhu02}Zhu, Y.; Granick, S. {\it Phys. Rev. Lett.}
{\bf 2002}, {\it 88}, 106102.

\bibitem{Cra01}Craig, V. S. J.; Neto, C.; Williams, D. R. M.
{\it Phys. Rev. Lett.} {\bf 2001}, {\it 87}, 054504.

\bibitem{Lau05}Lauga, E.; Brenner, M. P.; Stone, H. A. In \textit{Handbook of Experimental Fluid
Dynamics}; Foss, J., Tropea, C., Yarin, A., Eds.; Springer:
New-York (2005); cond-mat/0501557.

\bibitem{Net05}Neto, C.; Evans, D. R.; Bonaccurso, E.; Butt, H.-J.; Craig, V. S. J. {\it Rep. Prog. Phys.} {\bf 2005}, {\it 68}, 2859.

\bibitem{Pri04}Priezjev, N. V.; Troian, S. M. {\it Phys.
Rev. Lett.} {\bf 2004}, {\it 92}, 018302.

\bibitem{Bar99}Barrat, J.-L.; Bocquet, L. {\it Phys. Rev.
Lett.} {\bf 1999}, {\it 82}, 4671.

\bibitem{Cot03}Cottin-Bizonne, C.; Barrat, J.-L.; Bocquet,
L.; Charlaix, E. {\it Nature Materials} {\bf 2003}, {\it 2}, 237.

\bibitem{Jos06}Joseph, P.; Cottin-Bizonne, C.; Benoit, J.-M.; Ybert, C.; Journet, C.; Tabeling, P.; Bocquet,
L. {\it Phys. Rev. Lett.} {\bf 2006}, {\it 97}, 156104.

%\bibitem{Rei200}Reiter, G.; Khanna, R. {\it Phys. Rev. Lett.} {\bf 2000}, {\it 85}, 2753.

\bibitem{Was89}Wasserman, S. R.; Tao, Y.-T.; Whitesides, G. M. {\it
Langmuir} {\bf 1989}, {\it 5}, 1074.

%\bibitem{Jac998}Jacobs, K.; Herminghaus, S.; Mecke, K. R. {\it Langmuir} {\bf
%1998}, {\it 14}, 965.

%\bibitem{Bro96}Brochard-Wyart, F.; Gay, C.; de Gennes, P.-G. {\it
%Macromolecules} {\bf 1996}, {\it 29}, 377.

\bibitem{Rui99}de Ruijter, M. J.; De Coninck, J.; Oshanin, G. {\it Langmuir}
{\bf 1999}, {\it 15}, 2209.


\bibitem{Fet06}Fetzer, R.; Rauscher, M.; M\"unch, A.; Wagner, B. A.; Jacobs,
K. {\it Europhys. Lett.} {\bf 2006}, {\it 75}, 638.

\end{thebibliography}
\end{document}